\documentclass[twocolumn,prb,amsmath,amssymb]{revtex4}
\usepackage{graphicx}
\usepackage{dcolumn}
\usepackage{bm}
\usepackage {amssymb}
\usepackage {amsmath}

\begin{document}
\title{Negative differential resistance in single crystal
La$_{2}$CuO$_{4}$ at low temperature}

\author{B. I. Belevtsev}
\email{belevtsev@ilt.kharkov.ua} \affiliation{B. Verkin Institute
for Low Temperature Physics and Engineering, National Academy of
Sciences, pr. Lenina 47, Kharkov 61103, Ukraine}

\author{N. V. Dalakova}
\affiliation{B. Verkin Institute for Low Temperature Physics and
Engineering, National Academy of Sciences, pr. Lenina 47, Kharkov
61103, Ukraine}

\begin{abstract}
A current-controlled negative differential resistance has been
revealed in the $I$-$V$ characteristics of single crystal
La$_{2}$CuO$_{4+\delta}$ in the low temperature region. The
non-linear behavior of conductivity is accompanied by a transition
from positive to negative magnetoresistance when the current is
growing. Possible reasons for the effect observed are discussed.
\end{abstract}

\maketitle

La$_{2}$CuO$_{4+\delta}$ is a mother compound for one of the
family of high-$T_c$ superconductors (HTSC). The stoichiometric
La$_{2}$CuO$_{4}$ ($\delta = 0$) is an antiferromagnetic (AFM)
Mott insulator with the N\'{e}el temperature  $T_N\approx 320$~K.
On doping it with oxygen ($\delta \neq 0$), charge carriers
(oxygen holes) appear in the system \cite{kremer,tranq}, which
leads to destruction of the AFM order and brings the system into
the metallic (superconducting) state. The excess oxygen resides
between the LaO planes \cite{chai} and thus determines the
three-dimensional (3D) character of conductivity in
La$_{2}$CuO$_{4+\delta}$. Because of its high mobility, the excess
oxygen forms a favorable condition for chemical (impurity-induced)
phase separation. Indeed, as neutron-diffraction data show
\cite{jorg}, below 320 K crystalline La$_{2}$CuO$_{4+\delta}$
separates into two phases which are crystallographically close to
each other. One of the phases has stoichiometry similar to that of
La$_{2}$CuO$_{4}$. The other is rich in oxygen and becomes
superconducting below $T_c\approx 38$~K. More evidence supporting
phase separation in this material was obtained by quadrupole and
nuclear magnetic resonance techniques \cite{ueda}. The phase
separation and its investigation are topical problems of HTSC
physics. The structural and stoichiometric inhomogeneities caused
by phase separation can affect significantly the behavior of the
transport properties of copper oxides.
\par
This study is concerned with the effect of current upon the
conductive and magnetoresistive properties of single crystal
La$_{2}$CuO$_{4+\delta}$ ($T_N = 182$~K). The dc resistivity in
the direction parallel to the CuO$_2$ planes was measured using
the Montgomery method at different pre-assigned current. The
magnetic field was parallel to the tetragonal $\vec {c}$-axis of
the crystal and perpendicular to the transport current.

\begin{figure}[b]
\centering\includegraphics[width=0.9\linewidth]{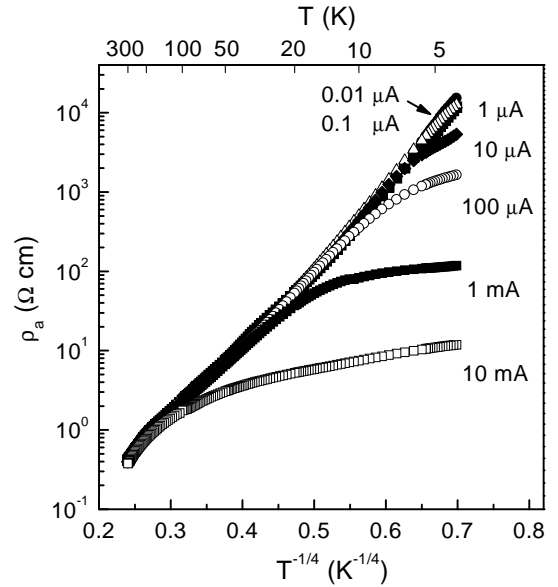}
\caption{The dependencies of $\lg \rho_{a}$ {\it vs.} $T^{-1/4}$
at different transport currents ($\vec{J}\parallel\vec{a}$)}
\end{figure}

\par
The temperature dependencies of the resistivity $\rho _{a}$
measured along the $\vec {a}$-axis for the different amplitudes of
the measuring currents $J$ are shown in Fig. 1. It is seen that at
$J \leq 1$~$\mu$A the resistance is only slightly dependent on
current in the whole interval of temperatures (4.2--300 K). The
Mott's law for variable-range hopping (VRH) is well obeyed in the
region 20--200 K for $J \leq 1$~$\mu$A:

\begin{equation}
\label{eq1} R \approx R_{0} \exp\left( {\frac{{T_{0}} }{{T}}}
\right)^{1/4},
\end{equation}
\noindent In this region the Ohm's law is obeyed well. The
exponent (1/4) in Eq. (\ref{eq1}) corresponds to the behavior of a
3D system.
\par
For $T < 10$~K and $J\leq 1$~$\mu$A the resistance grows with
lowering temperature more rapidly than it is predicted by Eq.
(\ref{eq1}). This behavior is typical for crystal
La$_{2}$CuO$_{4+\delta}$ in the low temperature region
\cite{boris1}. The effect may be produced by the isolated
superconducting inclusions that appear in the dielectric matrix on
phase separation when the volume fraction of the superconducting
phase is much smaller than the percolation threshold.

\begin{figure}[t]
\centering\includegraphics[width=0.9\linewidth]{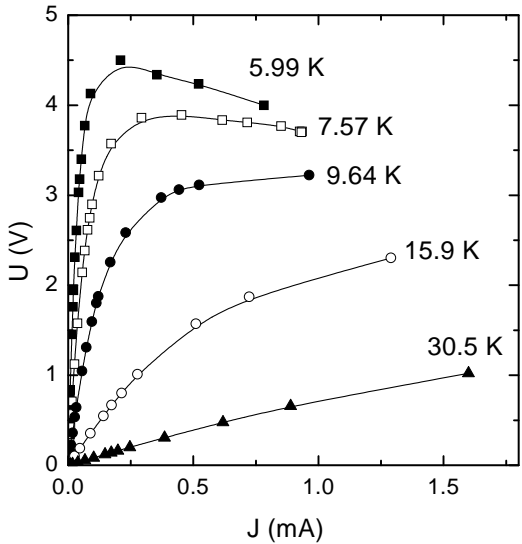}
\caption{$I-V$ characteristics for different temperatures.
\newline ($\vec {J}\parallel\vec{a}$)}
\end{figure}

\par
At $J > 1$~$\mu$A there is a significant deviation of
$\rho_{a}(T)$ from the Mott's law in low temperature region. When
the current increases, the resistance drops drastically, and the
temperature at which $\rho _{a}(T)$ starts to deviate from Eq.
(\ref{eq1}) shifts towards higher temperatures (Fig. 1). This
behavior accounts for the non-linear effects in the conductivity.
The non-linear $I$-$V$ curves are illustrated in Fig.~2. At $T <
8$~K some regions with negative differential resistance (NDR) can
be seen where $dV/dI < 0$. Earlier, a voltage-controlled NDR
effect at low temperatures ($T<10$~K) was observed in single
crystal La$_{2}$CuO$_{4}$ with inhomogeneous distribution of
oxygen \cite{boris2}. Here we report for the first time a
current-controlled NDR in single-crystal La$_{2}$CuO$_{4}$ with
more homogeneous distribution of oxygen.
\par
According to Ref. \onlinecite{mott}, the influence of electric
field on resistance under the VRH condition is described by

\begin{equation}
\label{eq2} R\left( {T,E} \right) = R_{0} \left( {T}
\right)\exp\left( { - \frac{{eEr_{h} \gamma} }{{kT}}} \right),
\end{equation}

\noindent where $R_{0}(T)$ is the resistance for $E \to 0$
described by Eq.~1, $r_{h}$ is the mean hopping distance, $\gamma$
is a factor of the order of unity. It is evident from Eq.
(\ref{eq2}) that in rather low fields ($E \ll kT/er_{h}\gamma$),
resistance is field independent, i.e the Ohm's law is obeyed. As
follows from estimation, this is true for the sample studied even
in the highest fields of the experiment. In this context the
non-linear behavior of the $I$-$V$ curves (Fig. 2) can hardly be
related to the influence of the electric field on hopping
conduction.

\begin{figure}[t]
\centering\includegraphics[width=0.9\linewidth]{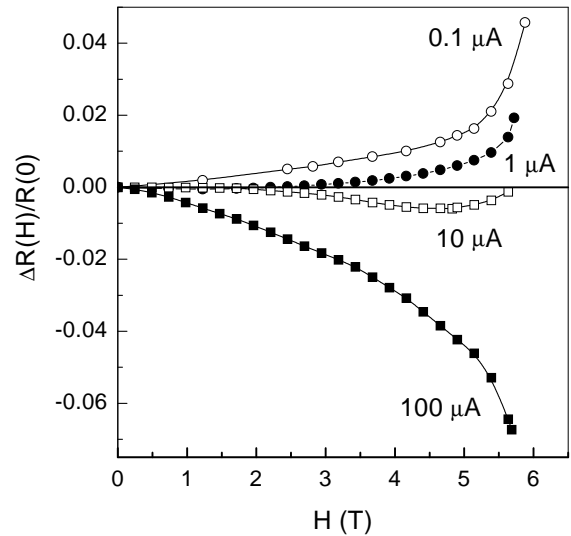}
\caption{Magnetoresistance curves taken at $T=5$~K for different
transport currents $J\parallel\vec{a}$ in the magnetic field $\vec
{H}\parallel\vec{c}$.}
\end{figure}

\par
There may be another reason for the non-linearity, namely,
electron overheating with rather high currents. If the charge
carriers do not have enough time to give up quickly the energy
received from the field to the lattice, their temperature rises
and exceeds that of the phonons. The overheating affects the
mobility of the carriers and leads to violation of the Ohm's law.
The theory of ``hot'' electrons was applied successfully to
explain the violation of the Ohm's law in experiments on doped
semiconductors \cite{con}. In Ref. \onlinecite{wang} the
non-linearity of experimental $I$-$V$ characteristics of doped Ge
with hopping conduction was described quantitatively taking into
account electron overheating and the ``thermal model'' of
electron-phonon energy transfer. It was assumed that the
resistance of the sample was determined only by the electron
temperature $T_{e}$ irrespective of the value of current.  In this
case the nonlinearity of $I$-$V$ curves was due to a decrease in
the sample resistance $R(T_{e})$ caused by the heating of the
charge carriers to $T_{e}$. As a result, the voltage over the
sample $V = IR(T_{e})$ can decrease when the current increases.
Below a certain critical temperature $T_{x}$ an extreme point
$dV/dI = 0$ appears in the $I$-$V$ curves, which is followed by a
NDR region. This is a region of instability, current and
resistance oscillations, and non-equilibrium transitions. The
known theories attribute NDR, among other things, to a non-uniform
distribution of impurities and defects over the crystal, which
produce regions with electric fields of different intensities. In
the sample studied, NDR can be caused by phase separation into
superconducting and dielectric regions.
\par
Qualitatively, the $I$-$V$ curves in Fig. 2 correspond to those
calculated in Ref. \onlinecite{wang} taking into account the
overheating effect. For the sample studied critical temperature
transition to NDR is about 6 K (Fig. 2); whereas estimations made
in the frame of the ``thermal model'' \cite{stef} give the value
close to 1 K. This discrepancy may be attributed with phase
separation into superconducting and dielectric regions. The model
in Refs. \onlinecite{wang,stef} was developed for semiconductors
and did not allow for superconducting inclusions as factors of
inhomogeneities. Nevertheless, the basic concepts of the model
\cite{wang,stef} account on the whole for the results obtained.
The observed current - controlled NDR effect can be interpreted as
NDR typical for percolation systems \cite{ridley} in which
increasing electric fields (currents) lead to elongation of the
existing high-conductivity percolation paths or even to the
formation of new ones. However, the results obtained are not
sufficient to analyze comprehensively or to draw conclusions about
particular mechanisms of this effect in the investigated sample.

\begin{figure}[htb]
\centering\includegraphics[width=0.9\linewidth]{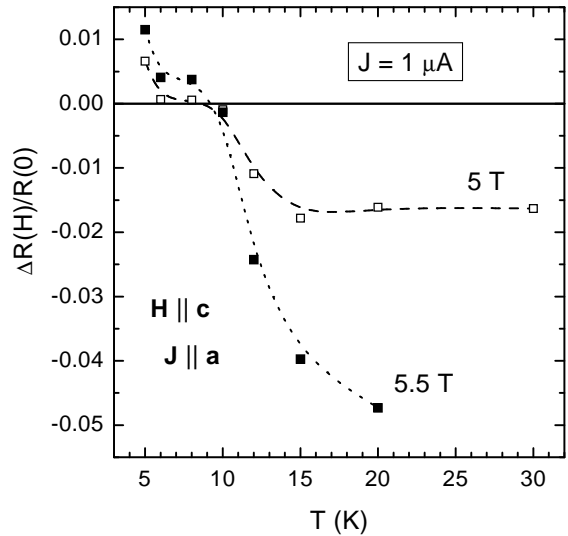}
\caption{Temperature dependencies of magnetoresistance for
$J\parallel\vec{a}$ taken at two magnitudes of magnetic field
$\vec {H}\parallel\vec{c}$.}
\end{figure}

\par
The behavior of magnetoresistance (MR) in the single crystal
La$_{2}$CuO$_{4+\delta}$ studied is also strongly dependent on
transport current and sensitive to electron overheating. The
effect of current is particularly evident in the low temperature
region (Fig. 3). For rather low currents $J \leq 1$~$\mu$A (with
conductivity close to the Ohmic one), MR is positive in the low
temperature region ($T \leq 10$~K) (Fig. 4). We can attribute this
positive MR to the influence of superconducting inclusions, like
in La$_{2}$CuO$_{4+\delta}$ sample with much higher $T_{N}$
\cite{boris2}. An increase in the current produces the Joule
heating and corresponding pair-breaking effect. As a result, the
positive MR disappears. When the current reaches $J \approx
10$~$\mu$A, MR becomes negative.
\par
The possible sources of the negative MR in
La$_{2}$CuO$_{4+\delta}$ at $T > 10$~K was considered in details
in Ref. \onlinecite{boris3}. In fields above $\approx 5$~T, MR is
to a large extent determined by the metamagnetic AFM - weak FM
transition. The competition of two different MR mechanisms and the
transition from positive to negative MR under electron overheating
are illustrated in Fig. 3. This corresponds to the temperature
behavior of MR at low currents (Fig. 4).
\par
The results of the MR investigation thus attest to the effect of
electron overheating, which in turn stimulates NDR at high
currents in the low temperature region. The latter effect evolves
from the inhomogeneous composition of the sample: because of phase
separation typical for this system, superconducting inclusions are
produced in the dielectric matrix at low temperatures.

\end{document}